\documentclass[10pt, conference, letterpaper]{IEEEtran}
\usepackage[letterpaper, left=0.63in, right=0.63in, bottom=1.01in, top=0.75in]{geometry}
\IEEEoverridecommandlockouts
\usepackage{cite}
\usepackage{amsmath,amssymb,amsfonts}
\usepackage{graphicx}
\usepackage[caption=false,font=footnotesize]{subfig}
\usepackage{textcomp}
\usepackage{xcolor}

\usepackage{tabularx}
\usepackage{multirow}
\usepackage{diagbox}
\usepackage[binary-units]{siunitx} 
\DeclareSIUnit{\bps}{bps}
\usepackage{amsthm}

\begin{document}

\title{Hierarchical Cell-Free  Massive MIMO for High Capacity with Simple Implementation}

\author{\IEEEauthorblockN{Wei Jiang\IEEEauthorrefmark{1} and Hans D. Schotten\IEEEauthorrefmark{2}}
\IEEEauthorblockA{\IEEEauthorrefmark{1}German Research Center for Artificial Intelligence (DFKI)\\Trippstadter Street 122,  Kaiserslautern, 67663 Germany\\
  }
\IEEEauthorblockA{\IEEEauthorrefmark{2}Rheinland-Pf\"alzische Technische Universit\"at  (RPTU) Kaiserslautern-Landau\\Building 11, Paul-Ehrlich Street, Kaiserslautern, 67663 Germany\\
 }
}
\maketitle

\begin{abstract}
Cell-free massive multi-input multi-output (MIMO) has recently gained much attention for its potential in shaping the landscape of sixth-generation (6G) wireless systems. This paper proposes a hierarchical network architecture tailored for cell-free massive MIMO, seamlessly integrating co-located and distributed antennas. A central base station (CBS), equipped with an antenna array, positions itself near the center of the coverage area, complemented by distributed access points spanning the periphery. The proposed architecture remarkably outperforms conventional cell-free networks, demonstrating superior sum throughput while maintaining a comparable worst-case per-user spectral efficiency. Meanwhile, the implementation cost associated with the fronthaul network is substantially diminished.
\end{abstract}

\section{Introduction}

Cell-free (CF) massive multi-input multi-output (MIMO) \cite{Ref_ngo2017cellfree} has recently garnered much attention in both academia and industry due to its high potential for sixth-generation (6G) systems\cite{Ref_jiang2021road}. There are no cells or cell boundaries. Instead, a multitude of distributed access points (APs) simultaneously serve a relatively smaller user population over the same time-frequency resource \cite{Ref_nayebi2017precoding}. It perfectly matches 6G private or campus networks, with relatively isolated coverage areas in scenarios like factories, stadiums, shopping malls, airports, railway stations,  exhibition halls, islands, or small towns. The CF architecture ensures uniform quality of service for all users, effectively addressing the issue of under-served areas commonly encountered at the edges of conventional cellular networks \cite{Ref_jiang2023cellfree}. Later, S. Buzzi \textit{et al.} proposed a \textit{user-centric} (UC) approach for CF massive MIMO \cite{Ref_buzzi2017cellfree, Ref_buzzi2020usercentric}, where each AP only serves a subset of users that are close to it.  UC can effectively lower the amount of fronthaul overhead while achieving comparable performance. 

Despite its considerable potential, CF still faces a lot of challenges, including the following two major concerns. Firstly, connecting a large number of distributed APs and a central processing unit (CPU) through a fronthaul network is costly \cite{Ref_masoumi2020performance}. Deploying a traditional wireless network is already arduous due to the complexities of acquiring and maintaining sites for base stations. In the CF architecture, the challenge is intensified as hundreds of suitable sites must be identified to accommodate wireless AP installations. The deployment of a massive-scale fiber-cable network to interconnect these APs further exacerbates the difficulty. In addition to the implementation cost, another concern revolves around uniform service quality, which is achieved at the price of system capacity degradation. Essentially, while the worst-case user rate is improved, the overall performance of other users is compromised through averaging.  Unlike the voice-oriented cellular networks like GSM in the 1990s, which demand uniform quality, the current 4G/5G networks, as well as the upcoming 6G systems, need to offer differentiated service quality tailored to the specific demands of diverse user devices and applications, rather than settling for averaged service \cite{Ref_jiang2021road}.

In this context, this paper proposes hierarchical cell-free (HCF) massive MIMO, an architecture that seamlessly integrates co-located and distributed antennas. A central base station (CBS), equipped with an antenna array, strategically positions itself at the heart of the coverage area, complemented by distributed APs spanning the periphery. The users are divided into two categories: near users (NUs) and far users (FUs). The NUs are connected to the CBS while each FU is served by a set of neighboring APs. In this way, the proposed HCF massive MIMO can offer the following advantages:
\begin{itemize}
    \item The implementation cost associated with the fronthaul network is substantially diminished because the service antennas located at the CBS do not need site acquisition and fiber connections. The CBS is dual-functional to replace the CPU in the conventional CF architecture. 
    \item Like the UC approach, the signaling overhead in terms of the number of complex-valued symbols exchanged in the fronthaul network is reduced since only a portion of APs close to each FU participate in communications.  
    \item The proposed architecture demonstrates superior sum throughput since the CBS offers reinforced service quality to the NUs. From the perspective of a user, its average data rate is improved accordingly.
    \item HCF maintains a comparable worst-case per-user rate measured by $5\%$-likely per-user spectral efficiency (SE). 
\end{itemize}

\begin{figure*}[!t]
    \centering
    \includegraphics[width=0.9\textwidth]{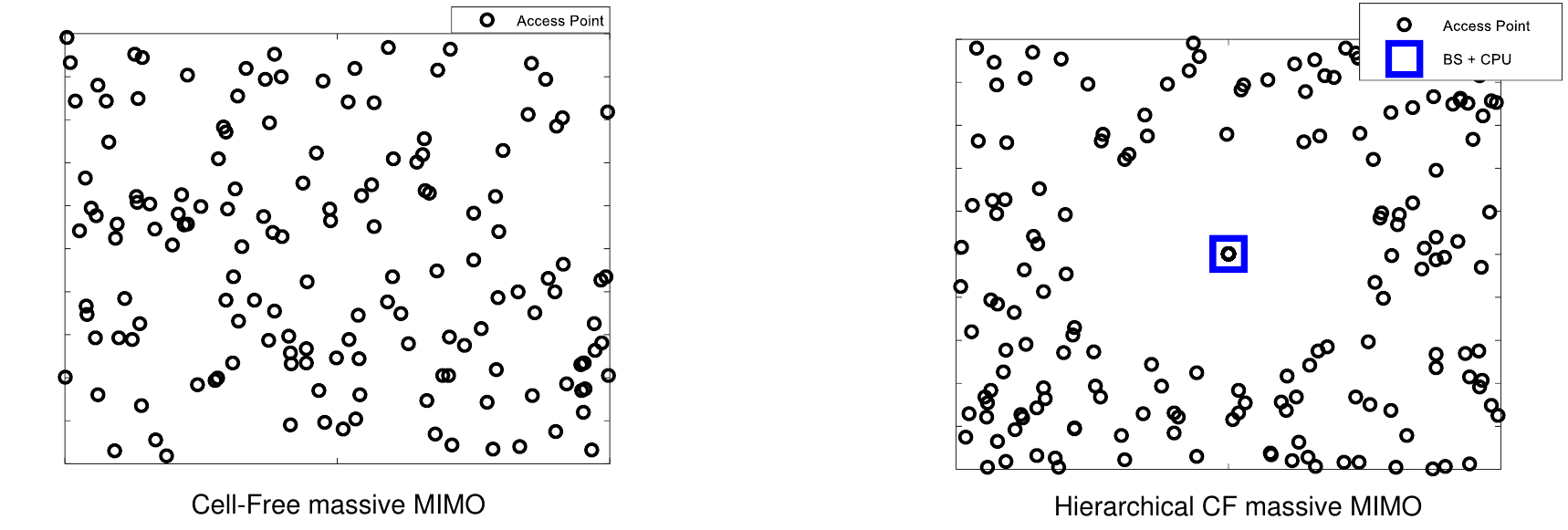}
    \caption{An illustrative comparison between the proposed hierarchical cell-free massive MIMO (right) and the conventional cell-free architecture (left).   }
    \label{fig:SystemModel}
\end{figure*}

\section{System Model}

In conventional cell-free massive MIMO, $M$ distributed APs serve a few $K\ll M$ user equipment (UEs) over an intended coverage area, as shown in the left one of \figurename \ref{fig:SystemModel}. Assume APs and UEs are equipped with a single antenna for simple analysis.  A CPU coordinates all APs through a fronthaul network. To avoid the prohibitive overhead of downlink pilots, which scales with the number of service antennas, time-division duplexing (TDD) is employed in massive MIMO to separate the downlink and uplink transmission. In the downlink, all APs transmit data symbols over the same time-frequency resource, while all UEs simultaneously send their signals in the uplink at another instant. 

In this paper, we propose a hierarchical architecture for CF massive MIMO, as shown in the right side of \figurename \ref{fig:SystemModel}, where a base station (BS) equipped with an array of $N_{b}$ antennas is located near the center of the coverage area. To differentiate the conventional BS, we name it central BS (CBS). It also functions as the CPU of the remaining $M-N_{b}$ distributed APs that are responsible for covering the edge area around the center. In this way, the cost of fronthauling is reduced since only a portion of service antennas is needed to be connected.  
The channel coefficient between antenna $m$, $\forall m=1,\ldots,M-N_b$ and UE $k$, $\forall k=1,\ldots,K$ is modeled as a circularly symmetric complex Gaussian random variable, i.e., $g_{mk}\in \mathcal{CN}(0,\beta_{mk})$, where $\beta_{mk}$ stands for large-scale fading including path loss and shadowing. The $N_{b}\times 1$ channel signature between the CBS and UE $k$, $\forall k=1,\ldots,K$ is denoted by $\mathbf{h}_k=[h_{1k},\ldots,h_{N_bk}]^T \in \mathcal{CN}(\mathbf{0},\beta_{k}^0\mathbf{I}_{N_b})$, where $\beta_{k}^0$ stands for the large-scale fading between the CBS and user $k$. Since large-scale fading is frequency-independent and varies slowly, the system measures it on a long-term basis and distributes it periodically. Thus, it is reasonable to assume that all nodes perfectly know this information.

In our HCF approach, the CBS first labels each user as a near user or a far user, according to a certain criterion, e.g., their distances to the CBS or their receiving signal strengths.  For instance, ordering the indices of the APs in terms of their large-scale fading in descending order, and then selecting some 'good' users to form a set of NUs $\mathbb{K}_0=\{k:\: \beta_k^0\geqslant \bar{\beta}_0\}$, where $\bar{\beta}_0$ is a pre-defined threshold for the CBS. The data symbols intended for the NUs are modulated and transmitted by the CBS in the downlink. In the uplink, all symbols from the NUs are detected while treating the FUs' signals simply as interference. On the other hand, the CBS determines a group of closest APs $\mathbb{M}_k=\{m:\: \beta_{mk}\geqslant \bar{\beta}_k\}$ to serve an FU $k\notin \mathbb{K}_0$, where $\bar{\beta}_k$ denotes the threshold for user $k$. From the perspective of a typical AP $m$, it maintains a list of associated users denoted by $\mathbb{K}_m=\{k:\: m\in\mathbb{M}_k \}$. In the downlink, the CBS only needs to deliver a portion of data symbols to a certain AP since an AP serves its closest users. This further lowers the overhead of fronthauling in comparison with the CF approach where each AP serves all users. In the uplink, each AP only processes the signals from the associated FUs $k\in \mathbb{K}_m$.

\section{The Communication Process}
Under the assumption of block fading, each coherent interval is divided into three phases: uplink training, uplink data transmission, and downlink data transmission. We ignore the time index of signals for simple analysis hereinafter.   

\subsection{Uplink Training}
During uplink training, UEs transmit orthogonal pilot sequences to acquire instantaneous channel state information (CSI). Unlike multi-cell systems, pilot contamination  \cite{Ref_zeng2021pilot} is avoidable by increasing the length of pilot sequences. Hence, we can neglect it for simplicity. A lot of literature like  \cite{Ref_ngo2017cellfree, Ref_nayebi2017precoding, Ref_buzzi2020usercentric, Ref_jiang2021cellfree, Ref_jiang2021impactcellfree} have already presented uplink training and therefore this paper does not repeat the details.  Conducting minimum mean-square error (MMSE) estimation, AP $m$ gets its local estimates $\hat{g}_{mk}\in \mathcal{CN}(0,\alpha_{mk})$, $\forall k=1,\ldots,K$ with $\alpha_{mk}=\frac{p_u\beta_{mk}^2}{p_u \beta_{mk} + \sigma_n^2}$, where $p_u$ and $\sigma_n^2$ denote the UE power constraint and the variance of additive noise, which is interchangeably denoted by $n$ or $w$ hereinafter. This estimation suffers from an error of $\tilde{g}_{mk}  = g_{mk} - \hat{g}_{mk}\in \mathcal{CN}(0,\beta_{mk}-\alpha_{mk})$.  Likewise, the CBS knows $\hat{\mathbf{h}}_{k}\in \mathcal{CN}(\mathbf{0},\alpha_{k}^0\mathbf{I}_{N_{b}})$, $\forall k=1,\ldots,K$ with $\alpha_{k}^0=\frac{p_u(\beta_{k}^0)^2}{p_u \beta_{k}^0 + \sigma_n^2}$ and the estimation error $\tilde{\mathbf{h}}_{k}  = \mathbf{h}_{k} - \hat{\mathbf{h}}_{k}\in \mathcal{CN}(\mathbf{0},(\beta_{k}^0-\alpha_{k}^0)\mathbf{I}_{N_{b}}) $,

\subsection{Uplink Data Transmission}
Because the UEs do not conduct channel estimation, data symbols are transmitted without channel-dependent phase offset. All UEs simultaneously send their signals towards the APs and CBS, where UE $k$ sends $x_k$ with a power coefficient $0\leqslant \eta_k \leqslant 1$. The covariance matrix of the transmit vector $\textbf{x}=[x_1,\ldots,x_K]^T$ satisfies $\mathbb{E}[\textbf{x}\textbf{x}^H]=\mathbf{I}_K$. The CBS observes 
\begin{equation} \nonumber
    \textbf{y}_b = \sqrt{p_u} \sum_{k=1}^K \sqrt{\eta_k} \mathbf{h}_k x_k + \mathbf{n}_b,
\end{equation} 
where the receiver noise $\textbf{n}_b\in \mathcal{CN}(\mathbf{0},\sigma^2_n\mathbf{I}_{N_b})$.
Aligning with \cite{Ref_ngo2017cellfree}, we apply matched filtering (MF), a.k.a. maximum-ratio combining, as the linear detector. It aims to amplify the desired signal as much as possible while disregarding inter-user interference (IUI). 
For each NU $k\in\mathbb{K}_0$, the CBS multiplies $\textbf{y}_b$ with $\hat{\mathbf{h}}_k^H$ to recover the transmitted symbol $x_k$, yielding a soft estimate of
\begin{align} \nonumber \label{GS_uplinksoftestimate}
    \check{x}_k &= \hat{\mathbf{h}}_k^H \biggl( \sqrt{p_u} \sum_{k=1}^K  \mathbf{h}_k \sqrt{\eta_k}x_k + \mathbf{n}_b \biggr)\\ \nonumber
    &=\underbrace{ \sqrt{p_u \eta_k} \| \hat{\mathbf{h}}_k\|^2  x_k }_{\mathcal{S}_0:\:desired\:signal} + \underbrace{ \sqrt{p_u \eta_k} \hat{\mathbf{h}}_k^H \tilde{\mathbf{h}}_k  x_k }_{\mathcal{I}_1:\:channel\:estimation\:error\:(CEE)}\\
    &+ \underbrace{\sqrt{p_u}\sum_{j=1,j\neq k}^K \hat{\mathbf{h}}_k^H \mathbf{h}_j \sqrt{\eta_{j}} x_{j}}_{\mathcal{I}_2:\:IUI}+\underbrace{\hat{\mathbf{h}}_k^H\textbf{n}_b}_{\mathcal{I}_3:\:noise},
\end{align}
applying $\textbf{h}_k=\tilde{\textbf{h}}_k+\hat{\textbf{h}}_k$.

Meanwhile, a typical AP $m$ observes 
\begin{equation} \label{GS_uplink_RxsignalAP}
    y_m = \sqrt{p_u} \sum_{k=1}^K \sqrt{\eta_k} g_{mk} x_k + n_m.
\end{equation} 
Similar to the UC approach \cite{Ref_buzzi2017cellfree}, the symbol $x_k$ from the FU $k\notin \mathbb{K}_0$ is merely processed on its associated APs $m\in\mathbb{M}_k$. That is to say, the $m^{th}$ AP, thus, form the statistics $\bar{y}_{mk}=\hat{g}_{mk}^*y_m$ for each $k\in \mathbb{K}_m$ and sends to the CBS. To detect $x_k$, the CBS generates a soft estimate $\check{x}_k = \sum_{m=1}^M \bar{y}_{mk}=\sum_{m\in \mathbb{M}_k} \hat{g}_{mk}^*y_m$. Utilizing \eqref{GS_uplink_RxsignalAP}, we have
\begin{align}  \nonumber \label{GS_ULsoftestAP}
    \check{x}_k &=\sum_{m\in \mathbb{M}_k} \hat{g}_{mk}^* \left( \sqrt{p_u} \sum_{j=1}^K \sqrt{\eta_j} g_{mj} x_j + n_m  \right) \\ \nonumber
     &=\sqrt{p_u\eta_k} \sum_{m\in \mathbb{M}_k} \| \hat{g}_{mk}\|^2 x_k +\sqrt{p_u\eta_k} \sum_{m\in \mathbb{M}_k} \hat{g}_{mk}^*   \tilde{g}_{mk} x_k  \\ 
     &+ \sqrt{p_u} \sum_{m\in \mathbb{M}_k} \hat{g}_{mk}^* \sum_{j=1,j\neq k}^K \sqrt{\eta_j} g_{mj} x_j +   \sum_{m\in \mathbb{M}_k} \hat{g}_{mk}^* n_m, 
\end{align}
with $g_{mk}=\hat{g}_{mk}+\tilde{g}_{mk}$.

\subsection{Downlink Data Transmission}
In the downlink, aligning with \cite{Ref_ngo2017cellfree}, conjugate beamforming (CBF) is applied to spatially multiplex the information-bearing symbols, i.e., $\textbf{u}=[u_1,\ldots,u_K]^T$, where  $\mathbb{E}[\textbf{u}\textbf{u}^H]=\mathbf{I}_K$. The CBS delivers a subset of symbols $\{u_k: k\in \mathbb{K}_m\}$ to the $m^{th}$ AP, resulting in lower fronthaul overhead than the CF approach that broadcasts all symbols $\{u_1,\ldots,u_K\}$ to every AP. Like the UC approach, AP $m$ transmits
\begin{equation} 
    s_m = \sqrt{p_d} \sum_{k\in \mathbb{K}_m} \sqrt{\eta_{mk}} \hat{g}_{mk}^* u_k,
\end{equation}
where $\eta_{mk}$ represents the power coefficient for the $k^{th}$ user at AP $m$, given per-antenna power constraint $p_d$.
Meanwhile, the CBS spatially multiplexes the information symbols intended for all NUs $\{k: k\in \mathbb{K}_0\}$. The transmitted signal at CBS antenna $n_b$ equals
\begin{equation} 
    d_{n_b} = \sqrt{p_d} \sum_{k\in \mathbb{K}_0} \sqrt{\eta_{n_bk}} \hat{h}_{n_bk}^* u_k,
\end{equation}
where $\eta_{n_b k_b}$ represents the power coefficient for the $k^{th}$ user at CBS antenna $n_b$.
As a consequence, a generic user $k$ has the observation of
\begin{align} \label{GS_DL_RxSig}
    y_k &=  \sum_{n_b=1}^{N_{b}} h_{n_bk} d_{n_b} + \sum_{m=1}^{M-N_{b}} g_{mk} s_m + w_k\\ \nonumber
      &= \sqrt{p_d} \sum_{n_b=1}^{N_{b}} h_{n_bk} \sum_{j\in \mathbb{K}_0} \sqrt{\eta_{n_bj}} \hat{h}_{n_bj}^* u_j \\ \nonumber
      &+ \sqrt{p_d}\sum_{m=1}^{M-N_{b}} g_{mk} \sum_{j\in \mathbb{K}_m} \sqrt{\eta_{mj}} \hat{g}_{mj}^* u_j + w_k.
\end{align}

\begin{figure*}[!t]
\setcounter{equation}{18}
\begin{align}\nonumber \label{GS_Downlink_RxSignalforNU}
    y_k  &= \underbrace{\sqrt{p_d}  \sum_{n_b=1}^{N_{b}} \sqrt{\eta_{n_bk}} \mathbb{E}\left[\| \hat{h}_{n_bk}  \|^2\right] u_k }_{desired\:signal}  +\underbrace{ \sqrt{p_d}  \sum_{n_b=1}^{N_{b}} \sqrt{\eta_{n_bk}}\left( \| \hat{h}_{n_bk}  \|^2-\mathbb{E}\left[\| \hat{h}_{n_bk}  \|^2\right]\right) u_k  }_{channel\:uncertainty\:error} + \underbrace{\sqrt{p_d}  \sum_{n_b=1}^{N_{b}} \sqrt{\eta_{n_bk}}\tilde{h}_{n_bk}   \hat{h}_{n_bk}^* u_k}_{channel\:estimation\:error} \\ 
    &+\underbrace{\sqrt{p_d} \sum_{n_b=1}^{N_{b}} h_{n_bk} \sum_{j\neq k, j\in \mathbb{K}_0} \sqrt{\eta_{n_bj}} \hat{h}_{n_bj}^* u_j}_{IUI\:from\:other\:NUs}+ \underbrace{\sqrt{p_d}\sum_{m=1}^{M-N_{b}} g_{mk} \sum_{j\in \mathbb{K}_m} \sqrt{\eta_{mj}} \hat{g}_{mj}^* u_j}_{IUI\:from\:FUs}+ w_k
\end{align} 
\rule{\textwidth}{0.1mm}
\begin{equation} \label{GS_SINR_DL_nearUser}
    \gamma_{nu,k}^{dl} =  \frac{  \left(\sum_{n_b=1}^{N_{b}} \sqrt{\eta_{n_bk}} \alpha_k^0 \right)^2   }
    {  \sum_{n_b=1}^{N_{b}} \beta_{k}^0 \sum_{j\in \mathbb{K}_0} \eta_{n_bj}  \alpha_j^0 + \sum_{m=1}^{M-N_{b}} \beta_{mk} \sum_{j\in \mathbb{K}_m} \eta_{mj} \alpha_{mj} + \sigma^2_n/p_d    }.
\end{equation}
\rule{\textwidth}{0.1mm}
\begin{align}\nonumber \label{GS_downlink_Rx_Signal_for_FU}
    y_k  &= \underbrace{ \sqrt{p_d}\sum_{m=1}^{M-N_{b}} \sqrt{\eta_{mk}} \mathbb{E}\left[\|g_{mk} \|^2\right] }_{desired\:signal}  +\underbrace{ \sqrt{p_d}\sum_{m=1}^{M-N_{b}} \sqrt{\eta_{mk}} \left( \|g_{mk} \|^2-\mathbb{E}\left[\|g_{mk} \|^2\right] \right)   }_{channel\:uncertainty\:error} + \underbrace{   \sqrt{p_d}\sum_{m=1}^{M-N_{b}} \sqrt{\eta_{mk}} \tilde{g}_{mk}  \hat{g}_{mk}^* u_k  }_{channel\:estimation\:error} \\ 
     &+ \underbrace{ \sqrt{p_d}\sum_{m=1}^{M-N_{b}} g_{mk} \sum_{j\neq k, j\in \mathbb{K}_m} \sqrt{\eta_{mj}} \hat{g}_{mj}^* u_j }_{IUI\:from\:other\:FUs}+\underbrace{ \sqrt{p_d} \sum_{n_b=1}^{N_{b}} h_{n_bk} \sum_{j\in \mathbb{K}_0} \sqrt{\eta_{n_bj}} \hat{h}_{n_bj}^* u_j    }_{IUI\:from\:NUs}  + w_k
\end{align}
\rule{\textwidth}{0.1mm}
\begin{equation} \label{GS_SINR_DL_FU}
    \gamma_{fu,k}^{dl} =  \frac{ \left( \sum_{m=1}^{M-N_{b}} \sqrt{\eta_{mk}} \alpha_{mk}\right)^2    }
    { \sum_{m=1}^{M-N_{b}} \beta_{mk} \sum_{j\in \mathbb{K}_m} \eta_{mj} \alpha_{mj}   + \sum_{n_b=1}^{N_{b}} \beta_k^0 \sum_{j\in \mathbb{K}_0} \eta_{n_bj}  \alpha_j^0 + \sigma^2_n/p_d    }.
\end{equation}
\rule{\textwidth}{0.2mm}
\setcounter{equation}{6}
\end{figure*}

\section{Performance Analysis}
This section analyzes the performance of the proposed HCF massive MIMO in terms of spectral efficiency. Per-user and sum SE in both downlink and uplink are provided.  
\subsection{Uplink Spectral Efficiency}
Distinct architectures of massive MIMO raise different levels of CSI availability. To be specific, the CBS has \textit{full CSI knowledge} of $\hat{\mathbf{h}}_k$,$\forall k$ as it receives the uplink pilots and conducts channel estimation. 
We derive the achievable SE for an NU $k\in \mathbb{K}_0$ as $R_{nu,k}^{ul}= \log(1+\gamma_{nu,k}^{ul})$, where the effective signal-to-interference-plus-noise ratio (SINR) equals
\begin{equation} \label{GS_UL_NU_SINR}
    \gamma_{nu,k}^{ul} =  \frac{  \eta_k N_b  \alpha_k^{0}   }{   \sum_{j=1}^K \eta_{j} \beta_j^{0}    -\eta_{k} \alpha_k^{0}   +  \frac{\sigma^2_n}{p_u}        }.
\end{equation}
\begin{IEEEproof}
The terms $\mathcal{S}_0$, $\mathcal{I}_1$, $\mathcal{I}_2$, and $\mathcal{I}_3$ in \eqref{GS_uplinksoftestimate} are mutually uncorrelated. According to \cite{Ref_hassibi2003howmuch}, the worst-case noise for mutual information is Gaussian additive noise with the variance equalling to the variance of $\mathcal{I}_1+\mathcal{I}_2+\mathcal{I}_3$. 
Thus, the achievable rate is lower bounded by $R= \log(1+\gamma)$,
where
\begin{align}  \label{cfmmimo:formularSNR}
    \gamma  = \frac{\mathbb{E}\left[|\mathcal{S}_0|^2\right]}{\mathbb{E}\left[|\mathcal{I}_1+\mathcal{I}_2+\mathcal{I}_3|^2\right]}
         = \frac{\mathbb{E}\left[|\mathcal{S}_0|^2\right]}{\mathbb{E}\left[|\mathcal{I}_1|^2\right]+\mathbb{E}\left[|\mathcal{I}_2|^2\right]+\mathbb{E}\left[|\mathcal{I}_3|^2\right]}
\end{align}
with 
\begin{align}  \label{APPEQ1}
    \mathbb{E}\left[|\mathcal{S}_0|^2\right] & = p_u\eta_k \left( N_b \alpha_k^{0} \right)^2\\ \label{APPEQ2}
    \mathbb{E}\left[|\mathcal{I}_1|^2\right] & = p_u\eta_k N_b \alpha_k^{0} (\beta_k^{0}-\alpha_k^{0})\\ \label{APPEQ3}
    \mathbb{E}\left[|\mathcal{I}_2|^2\right] & = p_u \sum_{j=1,j\neq k}^K \eta_j N_b \beta_j^{0} \alpha_k^{0}\\  \label{APPEQ4}
    \mathbb{E}\left[|\mathcal{I}_3|^2\right] & = \sigma_n^2N_b \alpha_k^{0}
\end{align}
Substituting these terms into \eqref{cfmmimo:formularSNR}, yields \eqref{GS_UL_NU_SINR}.
\end{IEEEproof}

In contrast, the CPU in the conventional CF architecture does not know CSI only if each AP delivers its local estimates or received uplink pilots to the CPU via the fronthaul network. However, it raises high signaling overhead. It is reasonable to assume that the CBS only knows the statistics of the channels between the users and APs. Consequently, the received signals are detected based on  $ \mathbb{E} \left[ \left |  \hat{g}_{mk} \right | ^2\right]=\alpha_{mk}$. Transform \eqref{GS_ULsoftestAP} into
\begin{align}  \nonumber \label{GS_ULsoftestAP_channelUncertainty}
    \check{x}_k  &=\underbrace{\sqrt{p_u\eta_k} \sum_{m\in \mathbb{M}_k} \mathbb{E}\left[\| \hat{g}_{mk}\|^2\right] x_k}_{\mathcal{S}_0:\:desired\:signal} +\underbrace{\sqrt{p_u\eta_k} \sum_{m\in \mathbb{M}_k} \hat{g}_{mk}^*   \tilde{g}_{mk} x_k}_{\mathcal{I}_1:\:CEE}  \\  \nonumber
     &+ \underbrace{\sqrt{p_u} \sum_{m\in \mathbb{M}_k} \hat{g}_{mk}^* \sum_{j=1,j\neq k}^K \sqrt{\eta_j} g_{mj} x_j}_{\mathcal{I}_2:\:IUI}+   \sum_{m\in \mathbb{M}_k} \hat{g}_{mk}^* n_m\\ & +\underbrace{\sqrt{p_u\eta_k} \sum_{m\in \mathbb{M}_k} \left( \| \hat{g}_{mk}\|^2 - \mathbb{E}\left[\| \hat{g}_{mk}\|^2\right] \right)x_k}_{\mathcal{I}_3:\:channel\:uncertainty\:error}, 
\end{align}
where an additional item $\mathcal{I}_3$ due to \textit{channel uncertainty} is imposed. 
The achievable SE for an FU $k\notin \mathbb{K}_0$ is $R_{fu,k}^{ul}= \log(1+\gamma_{fu,k}^{ul})$ with effective SINR of
\begin{equation} \label{GS_SINR_UL_AP}
    \gamma_{fu,k}^{ul} =  \frac{\eta_k \left( \sum_{m\in \mathbb{M}_k} \alpha_{mk}  \right)^2}
    {\sum_{m\in \mathbb{M}_k} \alpha_{mk} \sum_{j=1}^K \eta_j   \beta_{mj}  +  \frac{\sigma^2_n}{p_u} \sum_{m\in \mathbb{M}_k} \alpha_{mk}   }.
\end{equation} 
\begin{IEEEproof}
Likewise, in this case, we obtain
\begin{align} 
    \mathbb{E}\left[|\mathcal{S}_0|^2\right] & =  p_u \eta_k \left( \sum_{m\in \mathbb{M}_k} \alpha_{mk}  \right)^2  \\ 
    \mathbb{E}\left[|\mathcal{I}_1|^2\right] & =  p_u \eta_k  \sum_{m\in \mathbb{M}_k} \alpha_{mk} (\beta_{mk}-\alpha_{mk})  \\ 
    \mathbb{E}\left[|\mathcal{I}_2|^2\right] & = p_u \sum_{m\in \mathbb{M}_k} \alpha_{mk} \sum_{j=1,j\neq k}^K \eta_j \beta_{mj}
\end{align}
The loss caused by $\mathcal{I}_3$ is 
\begin{equation}
   \mathbb{E}\left[|\mathcal{I}_3|^2\right]  = p_u \eta_k\sum_{m\in \mathbb{M}_k} \alpha_{mk}^2  
\end{equation}
since 
\begin{equation}
    \mathbb{E}\left(\left|\| \hat{g}_{mk} \|^2 -\mathbb{E}[\| \hat{g}_{mk} \|^2]\right|^2\right)=\mathrm{Var}(\| \hat{g}_{mk} \|^2)=  \alpha_{mk}^2,
\end{equation}
Thus, we obtain the effective SINR as \eqref{GS_SINR_UL_AP}.
\end{IEEEproof}
The sum SE of the HCF massive MIMO system in the uplink is calculated by $ C_{ul}=\sum_{k\in \mathbb{K}_0} R_{nu,k}^{ul}+\sum_{k\notin \mathbb{K}_0} R_{fu,k}^{ul} $.

\subsection{Downlink Spectral Efficiency}
In the downlink, the $k^{th}$ user knows channel statistics $\alpha_{mk}$ or $\alpha_{k}^0$  rather than channel estimate $\hat{g}_{mk}$ or $\hat{\mathbf{h}}_{k}$ since there are no downlink pilots and channel estimation. As mentioned above, channel uncertainty error causes a loss since the received signals can only be detected using channel statistics. For an NU $k\in\mathbb{K}_0$, we need to further decompose \eqref{GS_DL_RxSig} into \eqref{GS_Downlink_RxSignalforNU} accordingly. Inter-user interference in this case consists of interference from other NUs and interference from the FUs. Due to the page limit, the detailed derivations of downlink SE are skipped.  Using similar manipulations as the derivation of uplink SE, we obtain the effective SINR as  \eqref{GS_SINR_DL_nearUser}. 
On the other hand, \eqref{GS_DL_RxSig} is rewritten to \eqref{GS_downlink_Rx_Signal_for_FU} from the FUs' perspective. 
Accordingly, we obtain the effective SINR of an FU $k\notin\mathbb{K}_0$ as \eqref{GS_SINR_DL_FU}. The downlink sum SE of the HCF massive MIMO system is computed by 
\begin{equation}
    C_{dl}=\sum_{k\in \mathbb{K}_0} \log(1+ \gamma_{nu,k}^{dl})+\sum_{k\notin \mathbb{K}_0}\log(1+ \gamma_{fu,k}^{dl}).
\end{equation}

\begin{figure}[!tbph]
    \centering
    \includegraphics[width=0.475\textwidth]{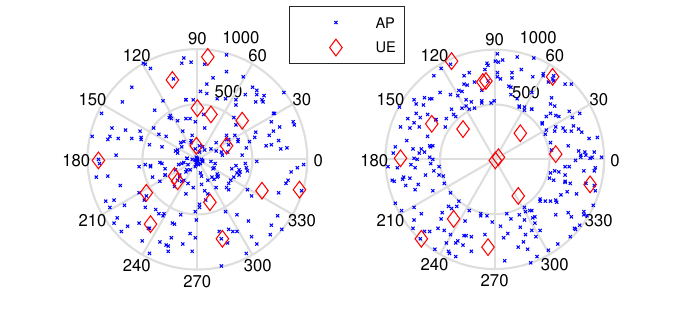}
    \caption{The simulation scenarios of CF in left and HCF in right. It shows the snapshot for a simulation epoch, where the locations of APs and UEs randomly vary in different epochs for ergodic evaluation.   }
    \label{fig:setup}
\end{figure}

\begin{figure*}[!tbph]
\centerline{
\subfloat[]{
\includegraphics[width=0.34\textwidth]{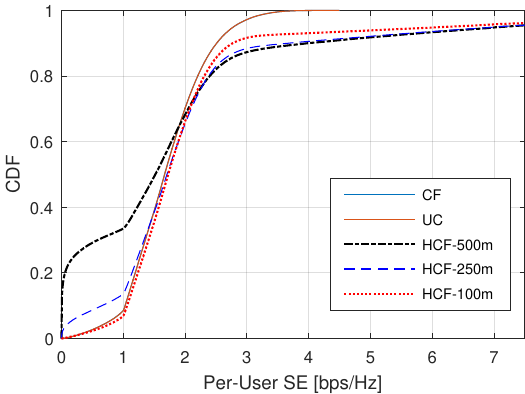}
\label{fig:result1} 
}
\hspace{1mm}
\subfloat[]{
\includegraphics[width=0.34\textwidth]{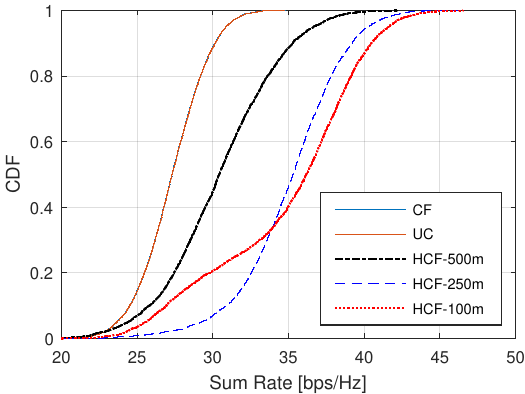}
\label{fig:result2} 
}
\hspace{1mm}
\subfloat[]{
\includegraphics[width=0.27\textwidth]{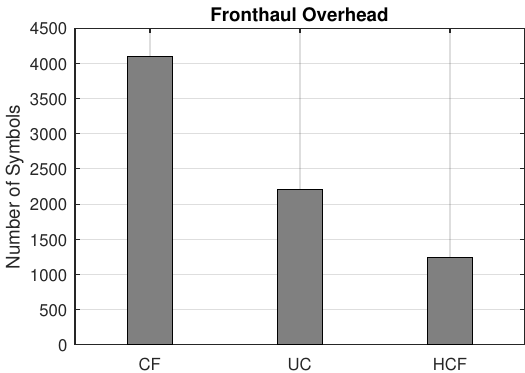}
\label{fig:result3}
}
}
\hspace{15mm}
 \caption{Performance comparison of CF/UC and HCF massive MIMO systems: (a) the CDF curves in terms of per-user SE; (b) the CDF curves of sum SE; and (c) the required number of complex-valued symbols delivered over the fronthaul network at each symbol period.    }
 \label{fig:result}
\end{figure*}

\section{Numerical Results}

The performance of the proposed HCF is numerically compared with that of CF and UC in terms of per-user SE, sum throughput, and fronthaul overhead. In our simulations, we establish a representative scenario where a total of $M=256$ antennas serve $K=16$ users within a circular area. In conventional CF/UC systems, $256$ APs and $16$ users are randomly distributed within a $1\mathrm{km}$ radius, as illustrated in the left diagram of \figurename \ref{fig:setup}. To implement the HCF architecture, we allocate half of the antennas to the CBS, i.e., $N_b=128$, aiming to minimize the scale of fronthaul networks as much as possible. This means that the HCF system in our simulations incurs only $50\%$ of the fronthauling implementation cost. Furthermore, $128$ single-antenna APs are distributed equally along the annulus with radii between $R=1\mathrm{km}$ and $r$, which takes three values — $100\mathrm{m}$, $250\mathrm{m}$, and $500\mathrm{m}$. The users falling into the inner circle are treated as NUs while the others are FUs. Varying $r$ allows us to observe the behavior of distant users at the cell edge in a traditional co-located antenna system. At each simulation epoch, the locations of APs and users randomly change, and a total of $10^5$ epochs are simulated for the ergodic performance.

Large-scale fading $\beta=10^\frac{\mathcal{L}+\mathcal{X}}{10}$, where the shadowing $\mathcal{X}\sim \mathcal{N}(0,\sigma_{sd}^2)$ with standard derivation $\sigma_{sd}=8\mathrm{dB}$, and the path loss is calculated by the COST-Hata model  \cite{Ref_ngo2017cellfree}:
\begin{equation} 
    \mathcal{L}= \begin{cases}
-L_0-35\log_{10}(d), &  d>d_1 \\
-L_0-10\log_{10}(d_1^{1.5}d^2), &  d_0<d\leq d_1 \\
-L_0-10\log_{10}(d_1^{1.5}d_0^2), &  d\leq d_0
\end{cases},
\end{equation}
where $d$ denotes the distance between a user and the CBS or an AP, the three-slope breakpoints  take values $d_0=10\mathrm{m}$ and $d_1=50\mathrm{m}$ while $L_0=140.72\mathrm{dB}$ in terms of 
\begin{IEEEeqnarray}{ll}
 L_0=46.3&+33.9\log_{10}\left(f_c\right)-13.82\log_{10}\left(h_{AP}\right)\\ \nonumber
 &-\left[1.1\log_{10}(f_c)-0.7\right]h_{UE}+1.56\log_{10}\left(f_c\right)-0.8
\end{IEEEeqnarray}
with carrier frequency $f_c=1.9\mathrm{GHz}$, the height of AP antenna $h_{AP}=15\mathrm{m}$, and the height of UE $h_{UE}=1.65\mathrm{m}$. Per-antenna and UE power constraints are set to $p_d=200\mathrm{mW}$ and $p_u=100\mathrm{mW}$, respectively.  The white noise power density equals $-174\mathrm{dBm/Hz}$ with a noise figure of $9\mathrm{dB}$, and the signal bandwidth is set to $5\mathrm{MHz}$. The uplink transmission is carried out in a distributed manner, it is reasonable that each UE simply use a full-power strategy $\eta_k=1$, $\forall k$ without global power control.

\figurename \ref{fig:result}a and \figurename \ref{fig:result}b compare the cumulative distribution functions (CDFs) of per-user SE and sum throughput, respectively, in the uplink data transmission. We implement user selection in both UC and HCF approaches on a per-user basis. That is, the threshold of user $k$ is calculated by $\bar{\beta}_k=\frac{1}{M}\sum_{m=1}^M \beta_{mk}$ and then build a group of closest APs $\mathbb{M}_k=\{m:\: \beta_{mk}\geqslant \bar{\beta}_k\}$. During the simulations, our observation is that the system performance does not affect too much about the exact values of $\bar{\beta}_k$. The key point is to exclude the users with the worst channel conditions, which causes performance degradation. In contrast, selecting or excluding a user with a moderate channel condition does not affect the performance. Under this user selection method, UC achieves identical per-user and sum SE as CF, and therefore their curves are completely overlapped in both figures. But UC has the advantage of reducing the fronthaul overhead, as shown in \figurename \ref{fig:result}c, resulting in approximately half of the data amount relative to CF.  

When the inner circle is big, namely $r=500\mathrm{m}$, some users far away from the CBS suffer from worse SE.  
To be specific, the $5\%$-likely per-user SE for HCF is close to zero, in comparison with $0.74\mathrm{bps/Hz}$ of CF and UC. However, HCF outperforms remarkably CF/UC since some users close to the CBS enjoyed the strengthened services, resulting in an average sum throughput of $30.6\mathrm{bps/Hz}$, which is better than $27.3\mathrm{bps/Hz}$ of CF/UC. If we shrink the inner circle to $r=250\mathrm{m}$, as  expected, the $5\%$-likely per-user SE can be improved to $0.14\mathrm{bps/Hz}$, while the average sum rate is increased to $35.1\mathrm{bps/Hz}$ accordingly. Further reducing the inner circle to $r=100\mathrm{m}$, it is amazing that the $5\%$-likely per-user SE of HCF even surpasses that of CF/UC, reaching $0.84\mathrm{bps/Hz}$. By far, we can conclude that the HCF architecture substantially improves the sum rate and per-user average rate, while it remains a comparable worst-case per-user rate if the network is properly configured in terms of $r$. Last but not least, we should keep in mind that the SE performance gain of HCF is achieved under a smaller scale fronthaul network, where only a half number of distributed APs are applied. Accordingly, approximately $50\%$ fronthaul overhead is further reduced compared to UC.

\section{Conclusion}
In this paper, we proposed a hierarchical network architecture tailored for cell-free massive MIMO, seamlessly integrating co-located and distributed antennas. A central base station equipped with an antenna array strategically positions itself at the heart of the coverage area, complemented by distributed access points spanning the periphery. Numerical evaluation justified that the proposed architecture remarkably outperforms conventional cell-free networks, demonstrating superior sum rates while maintaining comparable worst-case per-user rates. Meanwhile, the implementation cost associated with the fronthaul network is substantially diminished, adding a layer of economic viability to its technological advancements.

\bibliographystyle{IEEEtran}
\bibliography{IEEEabrv,Ref_COML}

\end{document}